\documentclass[sigconf]{acmart}

\usepackage{booktabs} 
\setcopyright{acmcopyright}

\copyrightyear{2017} 
\acmYear{2017} 
\setcopyright{acmcopyright}
\acmConference{SIGIR '17}{August 07-11, 2017}{Shinjuku, Tokyo, Japan}\acmPrice{15.00}\acmDOI{http://dx.doi.org/10.1145/3077136.3080759}
\acmISBN{978-1-4503-5022-8/17/08}

\begin{document}
\title{\huge Social Media Advertisement Outreach: Learning the Role of Aesthetics}

\author{Avikalp Srivastava}
\affiliation{%
  \institution{IIT Kharagpur}
}
\email{avikalp22@iitkgp.ac.in}

\author{Madhav Datt}
\affiliation{%
  \institution{Harvard University}
}
\email{madhav_datt@college.harvard.edu}

\author{Jaikrishna Chaparala}
\affiliation{
  \institution{IIT Kharagpur}
  }
\email{jaikrishna.ch@iitkgp.ac.in}

\author{Shubham Mangla}
\affiliation{%
  \institution{IIT Kharagpur}
  }
\email{shubhammangla@iitkgp.ac.in}

\author{Priyadarshi Patnaik}
\affiliation{%
  \institution{IIT Kharagpur}
  }
\email{bapi@hss.iitkgp.ernet.in}

\begin{abstract}
Corporations spend millions of dollars on developing creative image-based promotional content to advertise to their user-base on platforms like Twitter. Our paper is an initial study, where we propose a novel method to evaluate and improve outreach of promotional images from corporations on Twitter, based purely on their describable aesthetic attributes. Existing works in aesthetic based image analysis exclusively focus on the attributes of digital photographs, and are not applicable to advertisements due to the influences of inherent content and context based biases on outreach.

Our paper identifies broad categories of biases affecting such images, describes a method for normalizing outreach scores to eliminate effects of those biases, which enables us to subsequently examine the effects of certain handcrafted describable aesthetic features on image outreach. Optimizing on the features resulting from this research is a simple method for corporations to complement their existing marketing strategy to gain significant improvement in user engagement on social media for promotional images.
\end{abstract}

%
%

\maketitle

\vspace{-0.14in}
\section{Introduction}

In an effort to reach out to their user base, corporations spend millions of dollars developing creative image-based promotional content for social media platforms such as Twitter. The ability of corporations to engage a large portion of their target audience has very direct monetary consequences for them. Because of their focus on sales and brand promotion, these images come with certain inherent content and context based biases beyond just aesthetic attributes that influence overall outreach. 

Most aesthetic focused advertisement outreach research and development is based on data from advertisement quality surveys. Conducting such surveys is an extremely resource intensive task. There has also been significant work in aesthetic image analysis \cite{datta2006studying, datta2008algorithmic, marchesotti2011assessing, murray2012ava} for predicting user ratings of digital photographs. However, these studies cannot be applied to promotional images on social media, because, social media user engagement of an image, unlike user ratings, are influenced by multiple factors beyond just image aesthetics. 

In our paper, we develop an engagement score for images on Twitter, identify such broad "biases" or factors, propose an automated method to identify their presence in images and learn a transformation on scores to eliminate the effects of such biases. We select a set of hand-crafted describable image aesthetic features and train our system to learn the relative significance and influence of each of these features on engagement. To ensure that the results of our research are actionable by graphic designers, we restrict our work to human understandable/describable features and do not use generic or learned deep features such as the ones used in \cite{marchesotti2011assessing}.

We also go on to show that low level features that work for digital photography (not accounting for biases) do not work for social media ad success. We use the model described in \cite{datta2006studying} for aesthetics of digital photographs as a baseline. Our method performs significantly better (71.8\% vs 57.5\%) than the baseline on classifying our dataset of 8000 promotional images on Twitter into successful and unsuccessful images. 

In the end, we describe the basic function and design of an automated interactive system based on the results obtained from our model. The system takes promotional images from corporations as input and provides human understandable/describable aesthetic attributes of the image that may be tuned (for example, increasing spatial smoothness of hue property by 14\%) by designers to obtain the most significant increase in engagement on Twitter.  

Through this paper, our key contribution is developing a method to deal with the bias related challenges associated with analyzing effects of aesthetic features on outreach of social media advertisement images. This elimination of bias to give comparable image scores based only on aesthetic attributes, across different images and pages on social media, opens possibilities for research in computational aesthetics around the social media advertising industry.

\vspace{-0.05in}
\section{Data Collection}
We build a data set of 8,000 image based promotional tweets by scraping Twitter profiles of 80 different corporations. These corporations are particularly active on Twitter and have between 36,000 ($@Toblerone$) to 13 million ($@PlayStation$) followers, and 3,000 ($@tictac$) to 753,000 ($@united$) tweets. We select these corporations from across 20 broad categories such as retail, fast food, automobiles etc. to account for the diversity in promotional image representation. We scrape such image based promotional tweets along with their likes count, retweets count, date and time, page followers, page tweets and tweet text from the Twitter API for each corporation page, in proportion to their total number of tweets.

\begin{figure*}
\caption{From left to right: No bias present, holiday season, animal presence, human presence, discounts, product launch}
    \includegraphics[width=\textwidth]{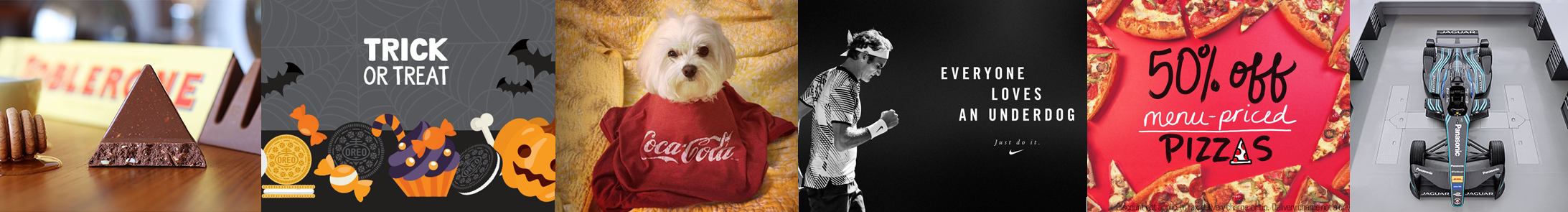}\hfill
\end{figure*}

For each such image $i$ $\in$ page $p$, we define our engagement evaluation score $\epsilon^i$, as the sum of image likes and image retweets. Due to inherent industry differences, and the variances across pages in total followers, we normalize our scores to ensure comparability between scores from different pages. We get a Pearson Correlation Coefficient of 0.46 and Spearman Rank Correlation of 0.63, suggesting no linear or monotonic correlation between number of followers and average engagement scores of a page. This is also supported by the sampled distribution in Figure 2. 

Therefore, in order to account for the difference in image based tweet engagement between pages of different sizes, industries etc. we use the mean and variance of engagement scores $(\epsilon^i)$ of images from the page $p$ for normalization. 
Thus, we define normalized engagement $\epsilon^i_N$: For image $i \in p$, $$\epsilon^i_N = \frac{\epsilon^i - \mu_p}{\sigma_p}$$ where $\mu_p$ and $\sigma_p$ are mean and standard deviation of image scores from page $p$.

\section{Bias driven engagement}

Datasets used by previous aesthetic image researches (Photo.Net \cite{datta2006studying}, DPChallenge \cite{datta2008algorithmic}, AVA \cite{murray2012ava}) only involve digital photographs rated by users purely based on their aesthetic appeal. Image based promotions on Twitter contain certain content and context based biases that significantly influence engagement scores $\epsilon_N$, for example, advertisements involving a cute cat will, on average, have much greater success and outreach, and consequently higher $\epsilon_N$ scores compared to aesthetically better images free from any significant biases. This makes analyzing the effects of aesthetic features on engagement an extremely challenging task. To ensure that our scores represent effects of, and are strongly correlated with visual appeal and aesthetic factors, we detect biases affecting each image and remove their influences on the score.

For each Twitter page within our 8000 image dataset, we detect outliers in terms of engagement scores using normalized local outlier factor scores and manually identify 8 broad categories of the most significant biases (listed in Table 1). In this paper, we account for and eliminate the effects of 4 biases - \textit{Animal Presence} (cats, dogs etc.), \textit{Human Presence} (babies, celebrities etc.), \textit{Special Days} (Black Friday, Christmas etc.), and \textit{Discounts}. Handling the remaining biases is beyond the scope of this paper and can give direction to future research in this area.

\begin{table}[!htbp]
\centering
\caption{Significant biases affecting Twitter engagement}
\begin{tabular}{|c|c|}
\hline
$\texttt{Discounts/Give-aways}$ & $\texttt{Human Presence}$   \\ 
\hline
$\texttt{Hashtags/Celebrity Mentions}$ & $\texttt{Animal Presence}$ \\
\hline
$\texttt{Special Days/Holiday Season}$ & $\texttt{Product Launches}$ \\
\hline
$\texttt{Social/Motivational Message}$ & $\texttt{Brand Popularity}$ \\     
\hline
\end{tabular}
\end{table}
\vspace{-0.10 in}
\subsection{Bias Identification}

We use the Viola-Jones face detector to give a binary classification detecting the presence of faces (as a proxy for human presence) in promotional images. To detect presence of animals, we train a spatial pyramid matching based SVM classifier described by Lazebnik et al. \cite{lazebnik2006beyond} on 5000 images of cats and dogs (most frequently occurring animals) scraped from the web. We manually identify all the 4 above mentioned biases from a sample of 1000 images from our 8000 image dataset, to assess quality of our automated bias identification, and obtained 75.5\% accuracy for human presence detection and 69.6\% accuracy for animal presence detection.

To extract text from the image, we use the Tesseract OCR Engine. In this paper, we define tweet text as the OCR extracted text along with the text associated with each image from its tweet. To account for the surge in Twitter engagement in periods leading up to major holidays such as Christmas, Black Friday, Halloween etc. we define date ranges around each such holiday (for example 7 days before and after Christmas). We manually build a list of $20$ words commonly associated with holidays (such as \texttt{Thanksgiving, Hanukkah} etc.) We augment this list of words by finding the $20$ most linguistically and semantically similar words using GloVe \cite{pennington2014glove}, which are then manually validated and filtered, and classify all tweets which occur within a holiday date range containing any of these words as Holiday biased. 

To identify tweets affected by biases caused by discounts/offers, we repeat the same process as described above using a different set of common initial words such as \texttt{free, discount, sale, offer} etc. with GloVe. We also identify tweets that urge users to retweet to get offers or win as part of some promotion. On our manually labeled 1000 image test set, we obtain 88.3\% accuracy on holiday-themed image identification and 84.4\% accuracy in identification of discounts and offers. While some images contain multiple biases, we restrict our paper to images with at most one bias which constitute a majority of our dataset (923 of our sampled 1000 images).

\begin{figure}
\caption{Distribution of Variances and Medians of Engagement Scores vs Number of Followers of Twitter Pages}    \includegraphics[height=2.0in,width=3.4in]{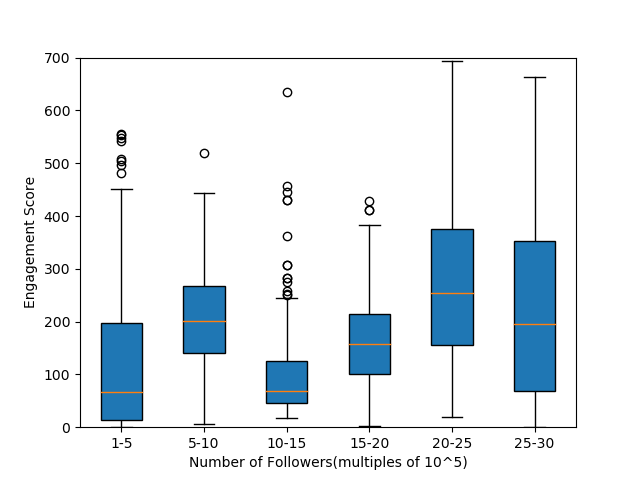}
\end{figure}
\begin{figure}
\caption{Left: Distribution of image scores with holiday bias. Right: Distribution comparison for unbiased, human-presence and animal-presence biased images. Both estimated via Gaussian curves}    \includegraphics[height=1.9in,width=3.4in]{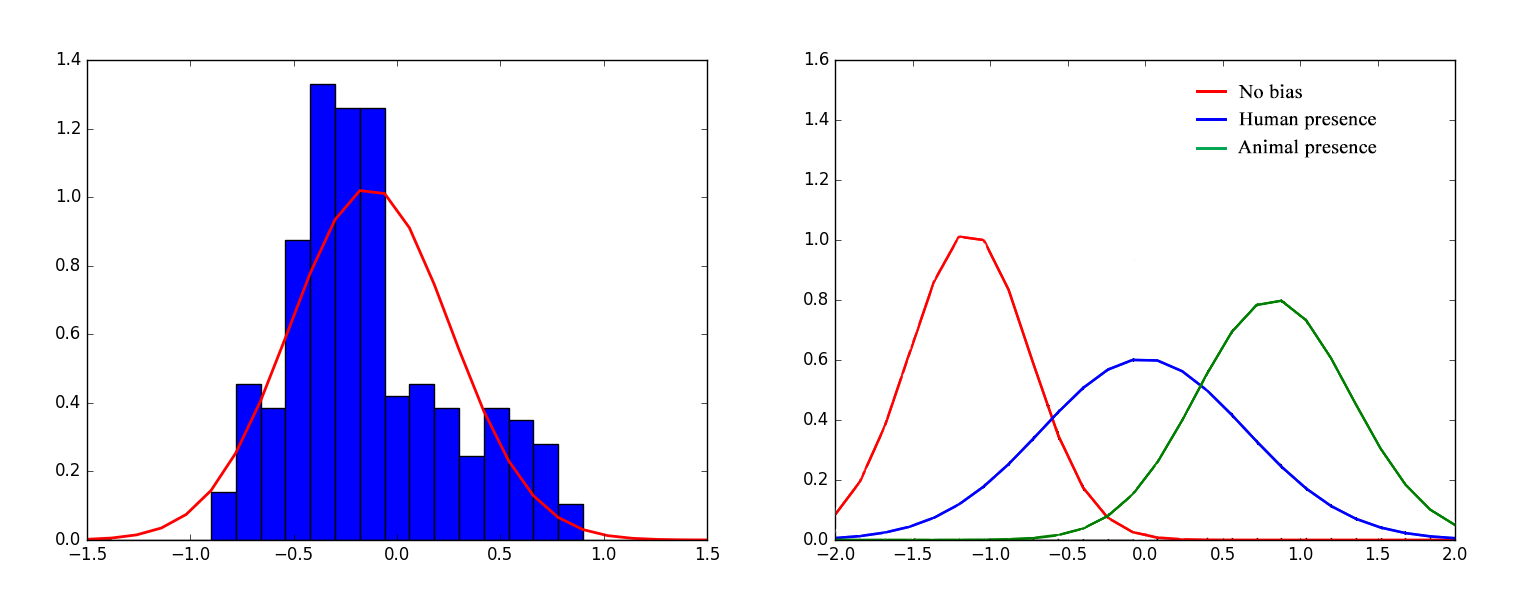}
\end{figure}
\vspace{-0.10 in}
\subsection{Bias Removal}
We define set of unbiased images $\mathbf{U}$ and mutually exclusive sets of identified biased images $\mathbf{B_j} \in \mathbf{B}$, where $\mathbf{B} = \{$Images with human presence, Images with animal presence, Holiday-themed images, Images with discounts$\}$. 

We define the discrete probability distribution of scores of images $\epsilon^i_N, \forall i \in \mathbf{U}$ as $\mathbf{P}$, and that of image scores $\epsilon^j_N, \forall j \in \mathbf{B_j}$, for $\mathbf{B_j} \in \mathbf{B}$, as $\mathbf{Q}$. To eliminate the effects of the bias, we apply the transformation  $\mathbf{Q} \xrightarrow{\tau} \mathbf{\bar{Q}}$, such that the distribution $\mathbf{\bar{Q}}$ can be used as an approximation of $\mathbf{P}$, while maintaining relative ranking order of image scores of biased images as in their original distribution $\mathbf{Q}$. Since, we transform $\mathbf{Q}$ to eliminate the effects of bias, $\mathbf{\bar{Q}}$ should be distributed similar to $\mathbf{P}$, and transformed image scores $\epsilon^i_{NT}$ for $i \in \mathbf{\bar{Q}}$ are used as if the original images had not been affected by the bias. To this effect, we use the Kullback-Leibler divergence $D_{KL}(\mathbf{P} || \mathbf{\bar{Q}})$ as our objective function for minimization, to capture the loss incurred while using $\mathbf{\bar{Q}}$ to approximate $\mathbf{P}$.

Let $Y_j$ be the set of scores for images in $\mathbf{B_j}$, thus the set of transformed scores $\bar{Y}_J$ is obtained as follows: $$\bar{Y_j} = \left\lbrace \bar{y_{ij}} ~ | ~ \bar{y_{ij}} = \tau(y_{ij}) \right\rbrace$$ where the probability distribution of the set $\bar{Y_j}$ is defined as $\mathbf{\bar{Q}_{\tau}}$. Thus, we learn the transformation $\tau^* \in \mathcal{D}$, where $\mathcal{D}$ is the space of all transformation functions, such that $$\tau^* = arg\min_{\tau \in \mathcal{D}} (D_{KL}(\mathbf{P} || \mathbf{\bar{Q}_{\tau}}))$$ \vspace{-0.005in} $$\textrm{where, } D_{\mathrm {KL} }(\mathbf{P} || \mathbf{{Q}})=\sum _{i \in Y_j}P(i)\,\log {\frac {P(i)}{Q(i)}}$$ In this paper, we explore the space of polynomial functions $D^p_d \in \mathcal{D}$, where the parameter $d$ defines the degree of polynomial for the input features $y_{ij}, \forall y_{ij} \in Y_j$. Therefore for the space $D^p_d$, $\tau$ as parameterized by $W = [W_0, W_1, .. W_d]$ on the input $y_{ij} \in Y_j$ is defined as follows: $$\tau_W(y_{ij}) = W_0 + W_1y_{ij} + W_2y^2_{ij} + .. + W_dy^d_{ij}$$ 
We add another constraint that $W \in ({\rm I\!R^+})^{(d+1)}$, so that during transformation, the relative ranking order of biased images is maintained and arbitrary transformations that disregard original distribution's ranking information and overfit are disallowed. For the bias associated with $Y_j$ in consideration, the input to the system is the matrix  $\mathbf{X} \in {\rm I\!R}^{(d+1) \times n}$ given by:
\[
\begin{bmatrix}
    1 & 1 & \dots & 1 \\
    y_{1j} & y_{2j} & \dots  & y_{nj} \\
    y_{1j}^2 & y_{2j}^2 & \dots  & y_{nj}^2 \\
    \vdots & \vdots & \ddots & \vdots \\
    y_{1j}^d & y_{2j}^d & \dots  & y_{nj}^d
\end{bmatrix}
\] where $n = |Y_j|$. At each step, the intermediate output transformation is given by $$\mathbf{O} = W^TX$$ where $\mathbf{O} = [y^`_{1j}, y^`_{2j}, ..., y^`_{nj}]$, where $y^`_{ij}$ represent the intermediate approximate values of the transformed score $\bar{y}_{ij}$. We minimize the above described divergence or loss function, $J(W) = D_{\mathrm {KL} }(\mathbf{P} || \mathbf{O})$ to learn the values for the matrix $W$. With the learned $W$, we apply the transformation $\tau^* (=\tau_W)$ to image scores in $Y_J$ to eliminate effects of bias described in $\mathbf{B_j}$. We repeat this process for each set of biased images in $\mathbf{B}$. 

\section{Aesthetic Feature Learning}
Computational assessment of aesthetic image quality is a well tackled problem. Having removed the content and context based biases associated with the scores received by the images in our dataset, we obtain image features strongly related with aesthetic appeal and visual attractiveness of the image to get the feature vector $ \mathbf{x}_i $ for each image \texttt{i}. We now formulate our feature learning problem with training data set $\lbrace \mathbf{x}_i, y_i \rbrace_{i\in[1,N]}$, where $y_i =  \epsilon^i_{NT}$, i.e. represents the transformed score of image $i$ after bias elimination. A function
$f : \mathbf{X} \rightarrow \mathbf{Y}$ is learned for providing feedback and suggestions for improving user engagement through image feature tuning.

\subsection{Feature Selection And Extraction}
We select describable/human-understandable image attributes based on handcrafted features used in previous works, augmenting the list with an additional set of features deemed important to capture the multi-object majority nature identified in advertisement images vis-a-vis photographic images. We implement the $56$ features defined by Datta et al. \cite{datta2006studying} in addition to non-overlapping features from Ke et al.\cite{ke2006design}, compositional attributes from Dhar et al.\cite{dhar2011high}, along with added features based on Region Adjacency Graphs (RAG) such as threshold and recursive normalized cut information. Thus, we obtain a total of $74$ describable aesthetic features for each image.

\begin{figure}
\caption{1st row: Images with highest feature values: Average hue for DoF, localized light exposure, average intensity of largest segment. 2nd row: Feature visualizations based on Region Adjacency Graphs}
    \includegraphics[height=1.5in,width=3.4in]{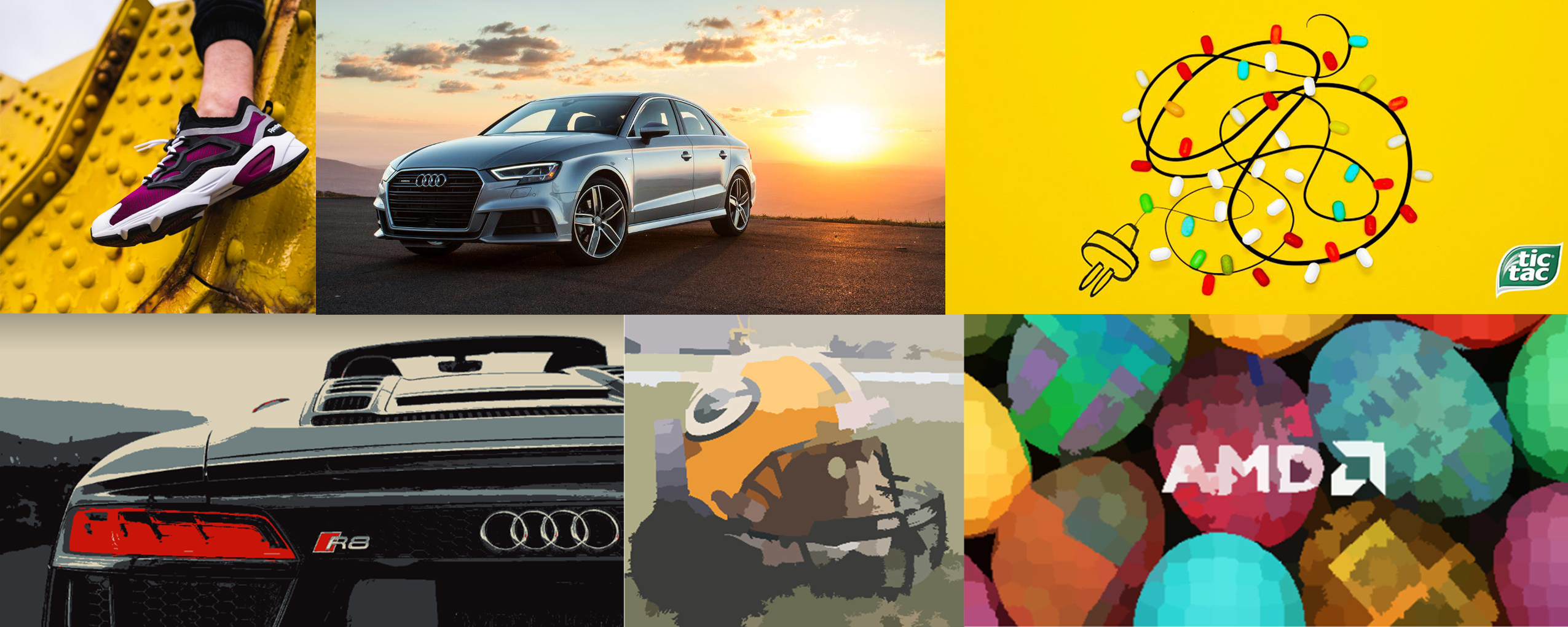}
\end{figure}

\subsection{Experimental Evaluation}
We use a standard support vector regressor with RBF kernel to learn the function $f : \mathbf{X} \rightarrow \mathbf{Y}$ where $\mathbf{X}$ denotes the $74$ feature based vector set and $\mathbf{Y}$ denotes the set of normalized and bias-removed engagement scores. This provides a quantitative evaluation of the relation between the feature vector values to the predicted engagement score, necessary for providing feedback on feature tuning for the query image to maximize the outreach through increased aesthetic and visual appeal, given that all other components for that image remain same. 

We also show that aesthetic features that work for digital photographs do not necessarily work for promotions/advertisement based images. From our dataset  $\lbrace \mathbf{x}_i, y_i \rbrace_{i\in[1,N]}$, we sample 20\% points to form our test set \textbf{T}.
To model our data for the classification task, we specify thresholds to partition images with scores in the lowest and highest quartile as "unsuccessful" and "successful" respectively.
We first run a SVM classifier model to learn weights for the $56$ features used in \cite{datta2006studying} on the Photo.net dataset and use this trained model to classify on our test set \textbf{T}.  While the accuracy of trained model on test set from Photo.net dataset is 69.12\% (close to value reported in \cite{datta2006studying}), on \textbf{T} this model's accuracy reduces to 57.5\%. Our 74 feature model trained on Twitter advertisement training dataset $\lbrace \mathbf{x}_i, y_i \rbrace_{i\in[1,N]} - \textbf{T}$ performs with 71.8\% accuracy on \textbf{T}, and on inspection we find that a good proportion of misclassified images contained biases we didn't handle in this study, and thus also good motivation for future work. The reduced accuracy achieved when using aesthetic features learned from non-advertisement dataset strongly suggests that it is necessary to capture image features linked with success of advertisement related images differently from those of purely aesthetically motivated digital photographs.


\begin{table}[!htbp]
\centering
\caption{The 5 highest significance attributes identified using linear kernel SVM 
for Photo.Net dataset vs. Twitter Advertisement Dataset}
\label{my-label}
\begin{tabular}{|c|c|}
\hline
\textbf{Photo.Net}                                                           & \textbf{Twitter Ads}                                                     \\ \hline
$\texttt{Familiarity measure}$                                                          & $\texttt{Low DoF hue component}$                                                    \\ \hline
$\texttt{Brightness measure}$                                                           & \begin{tabular}[c]{@{}c@{}}$\texttt{Largest segment}$\\ $\texttt{avg. intensity}$\end{tabular} \\ \hline
\begin{tabular}[c]{@{}c@{}}$\texttt{Avg. hue in wavelet}$\\ $\texttt{transformation}$\end{tabular} & $\texttt{Low DoF saturation}$                                                       \\ \hline
$\texttt{3rd largest patch size}$                                                       & $\texttt{RAG segment count}$                                                        \\ \hline
\end{tabular}
\end{table}

\begin{figure}
\caption{On input (a) to system with k=2, t=4\% suggestions: Increase features, light exposure by 24\%, spatial smoothness of 2nd level of saturation by 16\%; (b) final image with suggestions incorporated}    \includegraphics[height=1.8in,width=2.6in]{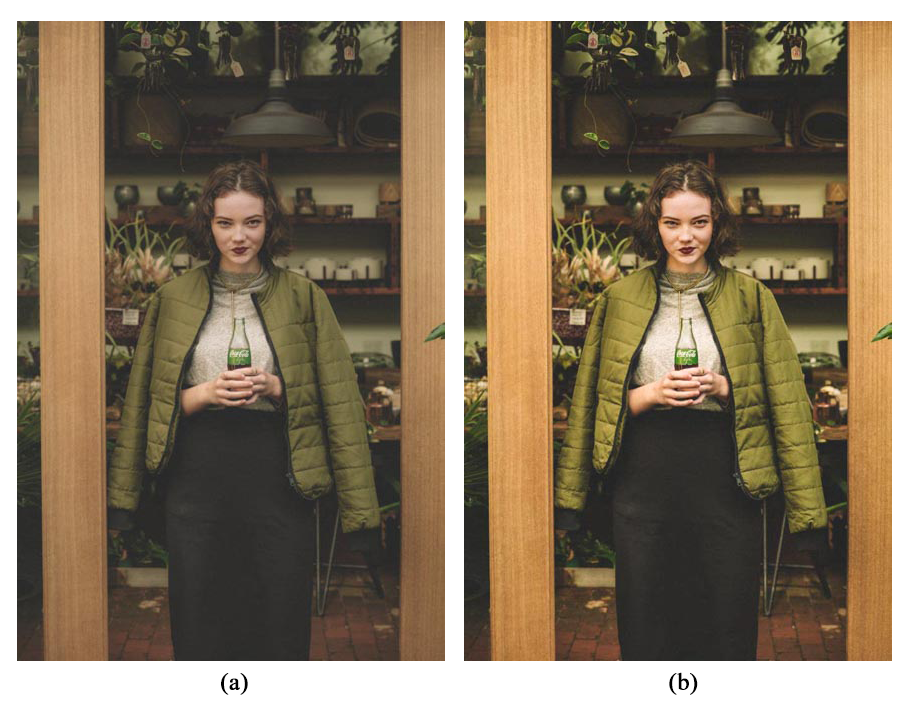}
\end{figure}

\vspace{-0.25cm}
\subsection{Applications and Feedback System}
Our paper describes the basic functioning and design of a system, based on our trained SVM regressor, to identify the aesthetic feature tunings that can be applied to image based promotions, in conjugation with other marketing strategies, to maximize user engagement.

Given an input promotional advertisement image \textit{i}, we apply our system for feature extraction to obtain the feature vector $\mathbf{x}_i$. Our system ideally seeks to find the nearest-neighbor feature vectors for $\mathbf{x}_i$ that lead to maximum increase in predicted engagement. That is, the system outputs a set of features and percentage changes for each, from our initially chosen human-understandable feature space, to maximize predicted engagement scores for the image. However, from a practical perspective, a graphics designer would be more interested in a tuning that provides changes to small number of features, rather than suggesting small changes to a large number of features, which can be inconvenient. A user may restrict the number of features where changes are suggested to at most \textit{k}, where each feature is not changed by more than a value \textit{s}. 
For finding the optimal tuning combination among these \textit{k} features, a distance of \textit{s} on either side of the original feature value is traversed in steps of size \textit{t}. The tuning combination that achieves the highest predicted engagement score from the trained support vector regressor is chosen as the suggestion output, as demonstrated in Figure 5. 



\section{Conclusion and Future Work} 
Our paper proposes a novel method to evaluate and improve outreach of promotional images from corporations on Twitter by identifying inherent biases and transforming scores to eliminate their effect on engagement in order to discover attributes that contribute most to advertisement outreach. Our model gives an aesthetic-feature based representation with corresponding outreach scores, enabling vector space model based retrieval strategies. It also opens new possibilities for research and applications of computational aesthetic analysis of images in the social media advertisement industry. Exploring and tackling the biases excluded by this study, using generic or deep learned features, computational improvements on the feedback system etc. promise exciting scope for future work.


\bibliographystyle{ACM-Reference-Format}
\small{
\bibliography{sigproc} 
}
\end{document}